\documentclass[%
 aip,
 amsmath,amssymb,
 reprint,%
]{revtex4-1}

\usepackage{graphicx}% Include figure files
\usepackage{dcolumn}% Align table columns on decimal point
\usepackage{bm}% bold math
\usepackage[utf8]{inputenc}
\usepackage[T1]{fontenc}
\usepackage{mathptmx}
\usepackage{amssymb,amsthm,amsmath,float,epsfig}
\usepackage{enumerate}
\usepackage{enumitem}
\usepackage{setspace}
\usepackage{cancel}
\usepackage{xcolor}

\theoremstyle{definition}

\newtheorem{Theorem}{Theorem}

\newtheorem*{Proof}{Proof}

\begin{document}

\preprint{AIP/123-QED}

\title[Geometric unfolding of synchronization dynamics on networks]{Geometric unfolding of synchronization dynamics on networks}.

\author{Llu\'is Arola-Fern\'andez}
\affiliation{Departament d'Enginyeria Inform\`{a}tica i Matem\`{a}tiques, Universitat Rovira i Virgili, 43007 Tarragona, Spain}%

\author{Per Sebastian Skardal}%
\affiliation{Department of Mathematics, Trinity College, Hartford, CT 06106, USA}%

\author{Alex Arenas}
\email{alexandre.arenas@urv.cat}
\affiliation{Departament d'Enginyeria Inform\`{a}tica i Matem\`{a}tiques, Universitat Rovira i Virgili, 43007 Tarragona, Spain}%

\date{\today}% It is always \today, today,

\begin{abstract}
We study the synchronized state in a population of network-coupled, heterogeneous oscillators. In particular, we show that the steady-state solution of the linearized dynamics may be written as a geometric series whose subsequent terms represent different spatial scales of the network. Namely, each addition term incorporates contributions from wider network neighborhoods. We prove that this geometric expansion converges for arbitrary frequency distributions and for both undirected and directed networks provided that the adjacency matrix is primitive. We also show that the error in the truncated series grows geometrically with the second largest eigenvalue of the normalized adjacency matrix, analogously to the rate of convergence to the stationary distribution of a random walk. Lastly, we derive a local approximation for the synchronized state by truncating the spatial series, at the first neighborhood term, to illustrate the practical advantages of our approach.

\end{abstract}

\maketitle

\begin{quotation}
Synchronization in ensembles of network-coupled heterogeneous oscillators plays a vital role in a number of natural and engineered phenomena, ranging from cell cycles to robust power systems~\cite{prindle2012sensing,witthaut12}. In a wide range of practical applications of such systems, including synchronization optimization~\cite{skardal14,skardal16a,taylor16}, control~\cite{skardal15b,skardal2016controlling}, and global entrainment~\cite{dorfler13}, the particular structure of the the fully synchronized state or a target synchronized state is critical for achieving the task at hand. In this paper, we study the synchronized state in networks of heterogeneous oscillators and show that it can written as a particular geometric series where subsequent series terms describe contributions from larger network neighborhoods. This spatial description of the state provides a contrast to, e.g., spectral methods that rely on global information. In particular, truncation of the geometric series at a fixed order is equivalent to building the synchronized state using only local, i.e., decentralized, information. We provide a rigorous proof of our main result, show that error in the truncated series decays geometrically with the second largest eigenvalue of the normalized adjacency matrix, and illustrate the practical advantages of this approach.
\end{quotation}

\section{Introduction}\label{section1}
The study of synchronization dynamics on networks of heterogeneous oscillators has shed light on a wide range of natural and engineered phenomena and has provided significant insights into structural properties of complex networks\cite{watts98,arenas08,pikovsky2003synchronization}. A particularly useful model for these purposes uses phase oscillators that evolve according to the following equations of motion,
\begin{equation}
\dot{\theta}_i = \omega_i + \sigma \sum_{j = 1}^N a_{ij}H(\theta_j-\theta_i),
\label{equation1}
\end{equation}
where $\theta_i$ is the phase of oscillator $i$, $i=1,\dots N$, $\omega_i$ is its natural frequency, $\sigma$ is the global coupling strength, and $H(\theta_j-\theta_i)$ is a $2\pi$-periodic non-linear coupling function of the state differences of pairs of connected oscillators. The network connectivity and the relative intensity of the interactions are captured by the entries $a_{ij}$ of the $N\times N$, real-valued adjacency matrix $A$ that in general can be undirected or directed, binary or weighted. While the particular choice of $H(\theta)$ depends on the application at hand, notable choices $H(\theta)=\sin(\theta)$ and $H(\theta)=\sin(\theta-\alpha)$, where $\alpha\in[-\pi/2,\pi/2]$, yield the network Kuramoto and Sakaguchi-Kuramoto models, respectively~\cite{kuramoto2003chemical}.

A particularly suitable framework useful for analyzing the interplay between structure and dynamics, described here by the network $A$ and the natural frequencies $\bm{\omega}$, respectively, emerges when a strongly synchronized, phase-locked state becomes attainable, for which we assume that $|H(0)|\ll1$, so that for a strongly synchronized state where $|\theta_j-\theta_i|$ we have that $H(\theta_j - \theta_i) \approx H(0) + H'(0)(\theta_j - \theta_i)$\cite{skardal14}. (We note here that for the classical Kuramoto model we have $H(0) = 0$ and $H'(0) = 1$.) By defining the effective frequency $\hat{\omega}_i = \omega_i + \sigma H(0)k_i^{in}$ and entering a suitable rotating frame, the synchronized state is then described by a the fixed point satisfying $\dot{\bm{\theta}} = 0$. After linearizing, the synchronized state then satisfies the following equation,
\begin{equation}
    \bm{\hat{\omega}} = \sigma H'(0) L \bm\theta,
    \label{equation2}
\end{equation}
where $L = D - A$ is the Laplacian of the network, and $D$ is a diagonal matrix with the in-degrees or in-strengths of the nodes, i.e. $D_{ii} = k_i = \sum_{j=1}^Na_{ij}$. Since the matrix ${L}$ has zero row sum, it has a trivial eigenvalue $\lambda_1 = 0$ with a constant associated eigenvector $\bm{v}_1\propto\bm{1}$, and therefore it is singular and not invertible\cite{newman10,skardal14}. Moreover, this spectral property reveals an important physical characteristic of the system, namely that the dynamics are invariant to a constant shift to the phases, i.e., translation along the synchronization manifold defined as the span of the trivial eigenvector $\bm{v}_1$. Thus, while solutions to Eq.~(\ref{equation2}) are not unique, that which minimizes the norm $||\bm{\theta}||$ is likely the most useful and is given by 
\begin{equation}
    \bm{\theta}^* = \frac{{L}^{\dagger} \bm{\hat{\omega}}}{\sigma H'(0)},
    \label{equation3}
\end{equation}
where ${L}^{\dagger}$ is the Moore-Penrose pseudo-inverse of the Laplacian matrix\cite{israel74}. Importantly, to write down the exact pseudo-inverse we require a full spectral decomposition of the Laplacian matrix (i.e., global information of the network). For a possibly directed network, the pseudo-inverse can be computed via the singular value decomposition of $L$ as \cite{skardal16a}
\begin{equation}
{L}^{\dagger} = \sum_{n = 2}^N \frac{\bm{v}_n \bm{u}_n^{T}}{\mu_n},
\label{equation4}
\end{equation}  
where $0 < \mu_2 \leq ... \leq \mu_N$ are the $N-1$ singular values of $L$ and $\{\bm{v}_n\}$ and $\{\bm{u}_n\}$ are the set of right and left singular vectors. Note that for the particular case of undirected networks the singular values are given by the eigenvalues of $L$ and the left and right singular vectors are are given by the eigenvectors of $L$, so $L^\dagger$ is defined byt its eigenvalue decomposition.

Eq.~(\ref{equation3}) gives precisely the linearized synchronized state and therefore is critical to understanding the interplay between structure and synchronization dynamics in tasks involving the fully synchronized state, and for this reason there is a myriad variety of phenomena and applications involving it. Eq.~(\ref{equation3}) emerges in the derivation of objective functions used to optimize the degree of synchronization in the system\cite{skardal14,skardal16a,skardal16b,taylor16} and in problems related to the control of desired synchronized states \cite{skardal15b}, critical points and in network inference techniques\cite{arenas06,timme07,dedomenico17}. Also, it is used to estimate analytically the local and global susceptibility of a power-grid system against structural failures\cite{manik17} and to rank specific nodes and edges according to its dynamical stability\cite{witthaut12,coletta16}. Furthermore, the object ${L}^{\dagger}$ provides, by its own, a lot of insight on the network properties in diffusive processes in general\cite{vanmieghem17}. In noisy dynamics, the coupling of ${L}^{\dagger}$ with the covariance matrix of the input noise determines the noise in the output signal such that the network can be optimized to perform as a noise-canceling filter\cite{Ronellenfitsch18}. It plays a central role in the theory of electric circuits, where it is used to obtain the resistance matrix that provides a notion of diffusion distance between nodes and to determine the best spreader of the network\cite{vanmieghem17}, as well as other node or edge centrality measures, such as the random-walk betweenness\cite{newman05}. From a mathematical point of view, ${L}^{\dagger}$ corresponds to the discrete Green's function of the network\cite{chung00} since it quantifies how a current injected in a node propagates to any other node of the network\cite{estrada08}.

All the previous results are expressed in the literature in terms of the spectra of $L$. The exact computation of ${L}^{\dagger}$ through the spectral or singular value decomposition can be computationally costly and the optimization and analysis of the interplay between structure and dynamics relies on numerical schemes that are often treated as black-boxes. Therefore, the networks studied in most of these cases are rather small and the interesting properties that emerge in the optimal coupling of topology and dynamics can only be observed \emph{a posteriori}, from the outcome of numerical methods. Also, the computations and interpretations of the results require global information of the network that is not available at the level of the nodes, which usually operate in a decentralized manner \cite{arenas08}.

Motivated by the previous limitations, we derive a geometric expansion of Eq.~(\ref{equation3}) in terms of increasingly further neighbourhoods of the nodes. This approach allows for a decentralized, fast and accurate computation of ${L}^{\dagger}\bm{\omega}$, up to a desired degree of accuracy (or amount of available information) and provides new analytical insights that allow for a deeper understanding of the interplay between topology and dynamics in network synchronization.

The remainder of this paper is organized as follows. In Sec.~\ref{section2} we present our main theoretical results, namely geometric series expansions for computing Eq.~(\ref{equation3}), along with convergence proofs. In Sec.~\ref{section3} we use local approximations, generated from our theoretical results to identify link removals that improve synchronization of networks. In Sec.~\ref{section4} we conclude with a discussion of our results.

\section{Main Results}\label{section2}

We begin by noting that rescaling the natural frequencies in Eq.~(\ref{equation2}) enables us to set $\sigma$ and $H'(0)$ both to one. To further simplify notation we drop the hat-notation, in which case Eq.~(\ref{equation2}) reduces to \cite{skardal14,skardal15b}
\begin{equation}
     \bm{\omega} = L \bm{\theta}.
     \label{equation5}
\end{equation}
To solve Eq.~(\ref{equation5}) without the Moore-Penrose pseudo-inverse we focus our attention more directly on $L$. First, using $L=D-A$, we write
\begin{equation}
    L = D(I-D^{-1}A).   
    \label{equation6}
\end{equation}
While $D$ is invertible (assuming that the network is connected and thus each node has some positive degree), $(I-D^{1}A)$ is not. This can bee seen by noting that $D^{-1}A$ is a stochastic matrix, and therefore has a leading eigenvalue $\lambda_1 = 1$. Then $I-D^{-1}A$ has a zero eigenvalue, making it singular. However, replacing $D^{-1}A$ with matrix $X$ that yields $(I-X)$ invertible, we have that 
\begin{equation}
    [D(I-X)]^{-1} = (I-X)^{-1}D^{-1}.
    \label{equation7}
\end{equation}
Moreover, the matrix $(I-X)^{-1}$ may be expanded in the geometric series
\begin{equation}
    (I-X)^{-1} = \sum_{m=0}^\infty X^m.
    \label{equation7a}
\end{equation}
The issue now arises that $X$ cannot be replaced by $D^{-1}A$, or more specifically, we have that
\begin{equation}
    [D(I-D^{-1}A)]^{-1} \ne \sum_{m=0}^\infty (D^{-1}A)^mD^{-1},
    \label{equation7b}
\end{equation}
namely, on the left hand side the inverse is ill-posed, and this is reflected by the fact that the series on the right-hand side diverges. However, this does not rule out the possibility of the right-hand side converging when it is applied to a vector of particular form. In fact, under the relatively mild conditions of the network having a primitive adjacency matrix, when the series is applied to an appropriately shifted frequency vector $\bm{\omega}$, the right-hand-side does converge and yields a solution to Eq.~(\ref{equation5}), which leads to the formulation of our first main result for undirected networks.
%The magnitude of the largest eigenvalue of the matrix $D^{-1}A$, i.e. the spectral radius $\rho(D^{-1}A)$ is exactly one because it is an stochastic row sum matrix \cite{israel74,chung00}. In principle, one cannot assure convergence of its geometric series in the form 
%\begin{equation}
%    (I-D^{-1}A)^{-1} \stackrel{?}{=} \sum\limits_{m=0}^\infty(D^{-1}A)^m,
%    \label{equation8}
%\end{equation}
%which shows why $L$ is not invertible in the first place \cite{israel74}. Remarkably, we find that, when proper conditions are met on the structure and the intrinsic dynamics of the system, the following result holds. 

\begin{Theorem}[Convergence of the geometric series for undirected networks]\label{theorem1}
Consider an undirected network with primitive adjacency matrix $A$ and a frequency vector $\bm{\omega}$ with zero mean, i.e., $\langle\bm{\omega}\rangle=0$.
%\cite{newman10,masuda17})
Then, the infinite series
\begin{equation}
\bm{\phi} = \sum_{m = 0}^{\infty} ({D}^{-1}{A})^m {D}^{-1} \bm{\omega},
\label{equation9}
\end{equation}
converges.
\end{Theorem}

\begin{Proof}
We begin by denoting the symmetric normalized adjacency matrix as $B=D^{-1/2}AD^{-1/2}$. Since $A$ is symmetric, so is $B$, and therefore its normalized eigenvectors $\{\bm{v}_j\}_{j=1}^N$ form an orthonormal basis for $\mathbb{R}^N$. Moreover, since $A$ is primitive, that is, there exists some integer $M>0$ such that $A^M$ is strictly positive (note that this is equivalent to $A$ being both irreducible and aperiodic), so is $B$. The Peron-Frobenius theorem~\cite{maccluer2000many} then implies that $B$ has a single largest eigenvalue $\lambda_1$ that is real and larger in magnitude than all other eigenvalues, i.e., $\lambda_1>|\lambda_j|$ for $j=2,\dots,N$. Moreover, since $B$ is normalized, we have $\lambda_1=1$ and $|\lambda_j|<1$ for $j=2,\dots,N$. Finally, the leading eigenvector associated with the leading eigenvalue $\lambda_1=1$ has entries that are proportional the the square root of the degrees of the respective nodes, i.e., $\bm{v}_1\propto\bm{k}^{1/2}$.

Next, it is useful to define
\begin{equation}
\bm{\phi}_m = {D}^{-\frac{1}{2}} {B}^m {D}^{-\frac{1}{2}}\bm{\omega},
\label{equation10}
\end{equation}
so that the right-hand-side of Eq.~(\ref{equation9}) is given by $\sum_{m=0}^\infty\bm{\phi}_m$. Defining the vector $\bm{x} = {D}^{-\frac{1}{2}}\bm{\omega}$, we now expand $\bm{x}$ via the orthonormal basis of eigenvectors of $B$, namely,
\begin{equation}
\bm{x} = \alpha_1 \bm{k}^{\frac{1}{2}} + \alpha_2 \bm{v}_2 + ... + \alpha_N \bm{v}_N,
\label{equation11}
\end{equation}
where $\alpha_i = \langle \bm{x}_i,\bm{v}_i \rangle$ are the coefficients given by projections of $X$ onto the different eigenvector directions and we assume that the eigenvector $\bm{v}_1 = \bm{k}^{1/2}$ is also appropriately normalized. Inserting Eq.~(\ref{equation11}) into Eq.~(\ref{equation10}) yields
\begin{equation} 
\begin{split}
{D}^{\frac{1}{2}}\bm{\phi}_m & ={B}^m \bm{x} \\
 & = \alpha_1 \lambda_1^m \bm{k}^{\frac{1}{2}} + \alpha_2 \lambda_2^m \bm{v}_2 + ... + \alpha_n \lambda_n^m \bm{v}_n.
 \label{equation12}
\end{split}
\end{equation}
Note now that for terms $j=2,\dots,N$, $\lambda_j^m$ decays geometrically while $\lambda_1^m=1$. However, we now show that the coefficient $\alpha_1$ must be zero. To see this, recall that the natural frequency vector has mean zero, or in other words, $\bm{\omega}$ is orthogonal to the constant vector $\bm{1}$, i.e., $\langle\bm{1},\bm{\omega}\rangle=0$. This is equivalent to $\langle D^{1/2}\bm{1},D^{-1/2}\bm{\omega}\rangle$, or more simply, $\langle\bm{k}^{1/2},\bm{x}\rangle=0$, which is precisely $\alpha_1$. Thus, we have that
\begin{equation} 
{D}^{\frac{1}{2}}\bm{\phi}_m  = \alpha_2 \lambda_2^m \bm{v}_2 + ... + \alpha_n \lambda_n^m \bm{v}_n.
 \label{equation12a}
\end{equation}
The convergence of the right-hand-side of Eq.~(\ref{equation9}) now follows quite easily: since each of the finitely-many series $\sum_{m=0}^\infty \alpha_j\lambda_j^m\bm{v}_j$ converges to $\alpha_j\bm{v}_j/(1-\lambda_j)$ for $j=2,\dots,N$, we have that the full series converges to
\begin{equation}
\begin{split}
\phi &= D^{-1/2}\sum_{m=0}^\infty D^{1/2}\phi_m\\
&= D^{-1/2}\sum_{m=0}^\infty \sum_{j=2}^N\alpha_j\lambda_j^m\bm{v}_j\\
&= D^{-1/2}\sum_{j=2}^N\frac{\alpha_j\bm{v}_j}{1-\lambda_j},
\label{equation12b}
\end{split}
\end{equation}
which concludes the proof. \qed
\end{Proof}

A subtle issue that arises with the result given in Theorem~\ref{theorem1}  is that, while the infinte series applied to the zero-mean vector $\bm{\omega}$ converges, it does not necessarily converge to the minimum norm solution given by Eq.~(\ref{equation3}) which turns out to the the zero-mean solution recovered by the Moore-Penrose pseudo-inverse approach. This can be seen by noting that at initial truncation of the geometric series we have that $\langle \bm{\phi}_0 \rangle = \langle \frac{\omega}{k} \rangle$, which is not zero. However, this can be easily fixed by simply applying a constant shift to the resulting vector to ensure that it is orthogonal to the constant eigenvector $\bm{v}_1$ and the solution converges to the one with minimal norm. In fact, this correction leads us to a generalization of our main result, which importantly applies to both undirected and directed networks.

\begin{Theorem}[Convergence of the geometric series: General case]\label{theorem2}
Consider a network with primitive adjacency matrix $A$. Then, the infinite series
\begin{align}
\bm{\phi}=\sum\limits_{m=0}^\infty \left(\bm{\phi}_m-\langle\bm{\phi}_m\rangle\right),\label{equation13}
\end{align}
where
\begin{align}
\bm{\phi}_m=(D^{-1}A)^mD^{-1}\bm{\omega},\label{equation14}
\end{align}
converges. 
\end{Theorem}

\begin{Proof}
The complete proof is shown in Appendix~\ref{appendixA}.\qed
\end{Proof}

\begin{figure}[ht]
\includegraphics[scale = 0.35]{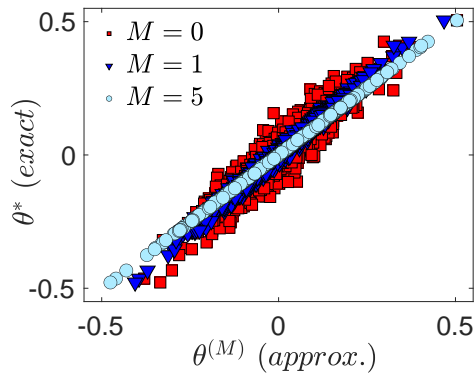}
\caption{\label{fig1} Scatter-plot of the exact Eq.(\ref{equation3}) vs the truncated Eq.(\ref{equation14}) approximation of the stationary phases, for different neighborhood orders (M), in a fixed Erd\"os-R\'enyi network of size $N = 500$, mean degree $\langle k \rangle = 20$ with a normal distribution of frequencies $N(0,1)$ for three different truncation orders.}
\end{figure}

Before proceeding to other practical questions and examples, we remark that Theorem~\ref{theorem2} includes no zero mean condition on the frequency vector nor is it restricted to undirected networks, thus its potential usage is wider than Theorem~\ref{theorem1}. In Fig.~(\ref{fig1}) we illustrate the utility of the more general case given by Eqs.~(\ref{equation13}) and (\ref{equation14}) when applied to a random network with random allocated frequencies. We truncate the expansion at different neighborhood orders $M$ and compare the approximated solution against the exact one given by Eq.(\ref{equation3}), showing that, for this particular random configuration, the approximation is accurate even for small $M$.

Investigating our results described in Eq.~(\ref{equation9}) and Eqs.~(\ref{equation13}) and (\ref{equation14}) more deeply, we see that the geometric expansion expresses the solution of the linear system as a sum of contributions of terms $\sim \omega/k$ coming from increasingly further neighborhoods of the nodes. That is, the $m^{\text{th}}$ order term consists of terms $\omega_j/k_j$ corresponding to oscillators located precisely $m$ links removed from a given oscillator. Thus, the series is simply the Taylor expansion of the solution expanded about each oscillator with higher-orders corresponding to larger network neighborhoods. It constructs the exact solution by adding infinitely many incremental pieces of local information in a polynomial basis that is not necessarily orthogonal. Interestingly, this differs from the spectral decomposition used for $L^\dagger$ in Eq.~(\ref{equation3}). In the latter, the solution is constructed by adding $N-1$ pieces of global information (i.e. rank-one matrices for each non-zero eigenvalue and its associated eigenvector in Eq.(\ref{equation4})), analogously to a Fourier expansion that expresses the solution in the basis of orthogonal eigenmodes with its associated eigenfrequencies (which carry global information on the original function)\cite{mcgraw07,shuman13}.

Furthermore, there is a clear connection between the applicability of this method and Markov chains and random walks \cite{masuda17}. In particular, for a Markov chain to have a unique, globally attracting stationary state it is necessary and sufficient to have a primitive transition matrix (or, equivalently, irreducible and aperiodic). For a network with adjacency matrix $A$ the transition matrix for a random walk is given by $D^{-1}A$ (or $AD^{-1}$, depending on the definition of the random walk). Here $D^{-1}A$ appears in the geometric series because the solution is expressed in terms of ``in-neighborhoods''. Since this ``spatial'' expansion unfolds in terms of in-neighborhoods of radius zero, one, two, three, etc., convergence of the expansion implies that eventually these neighborhoods must include the whole network, i.e., each node must be reachable from each node in a manner that is eventually well-mixed, hence the need for the network adjacency matrix to be irreducible and aperiodic, i.e., primitive\cite{masuda17}.

\begin{figure}[ht]
\includegraphics[scale = 0.153]{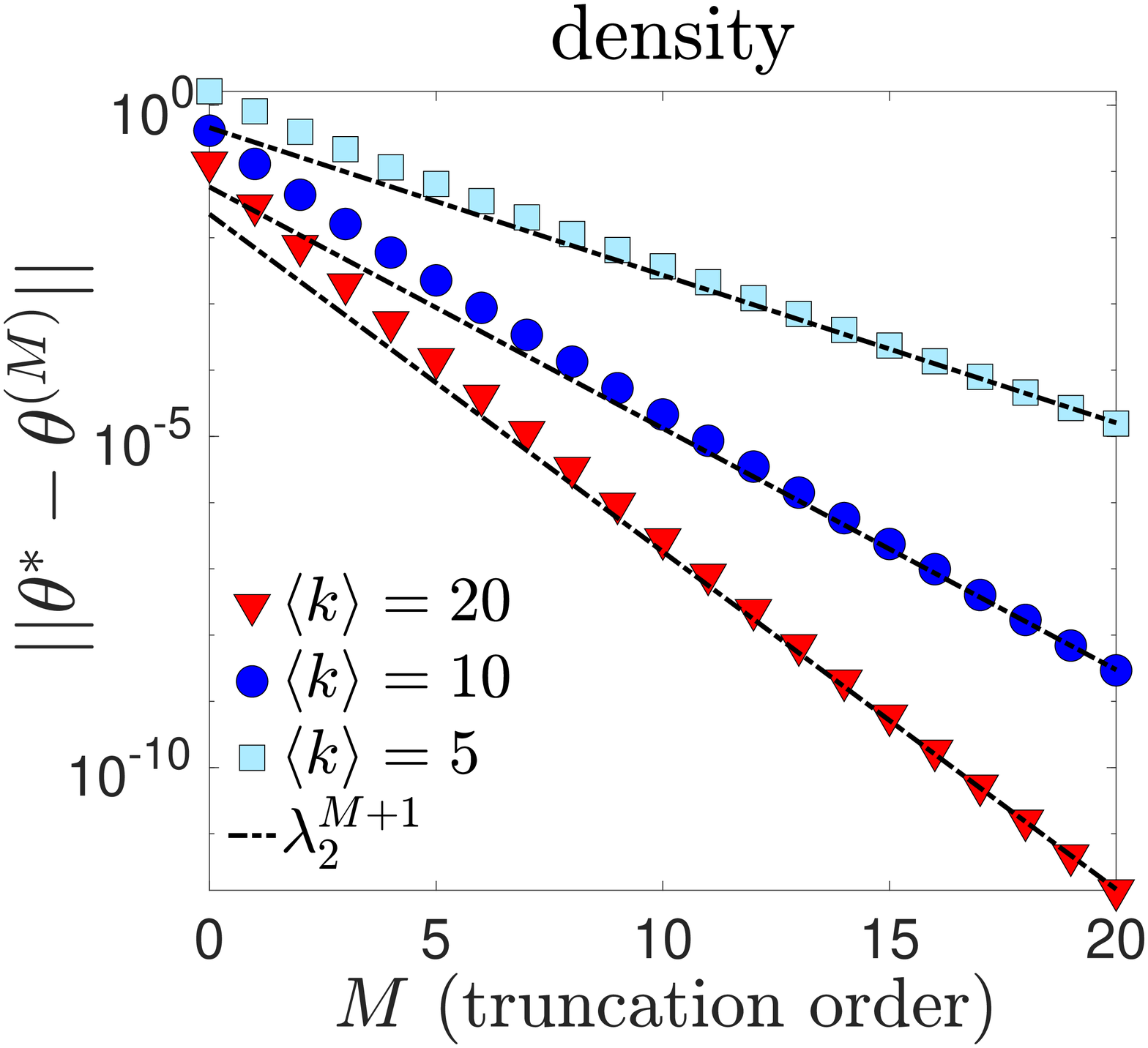}
\includegraphics[scale = 0.153]{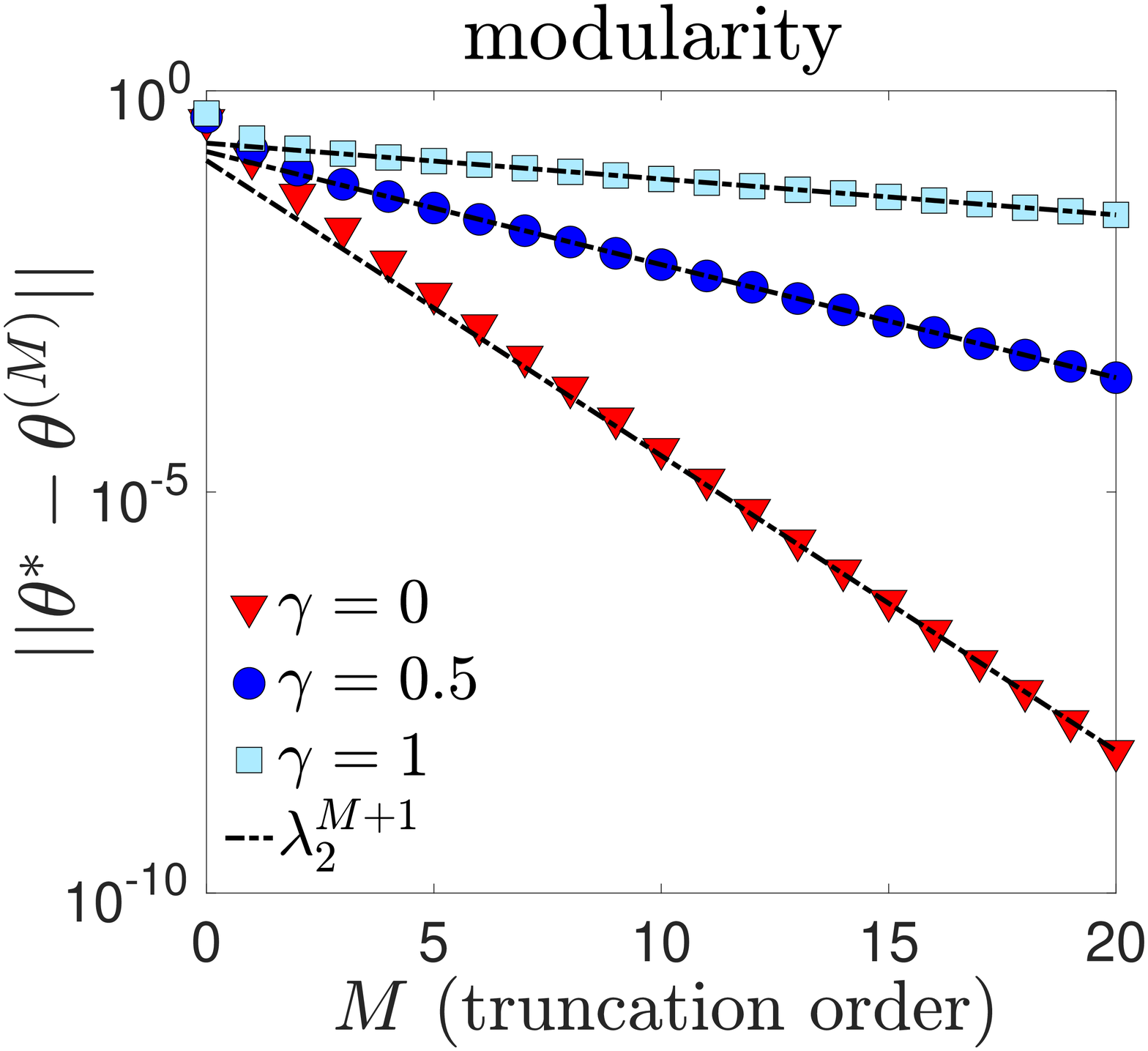}
\includegraphics[scale = 0.153]{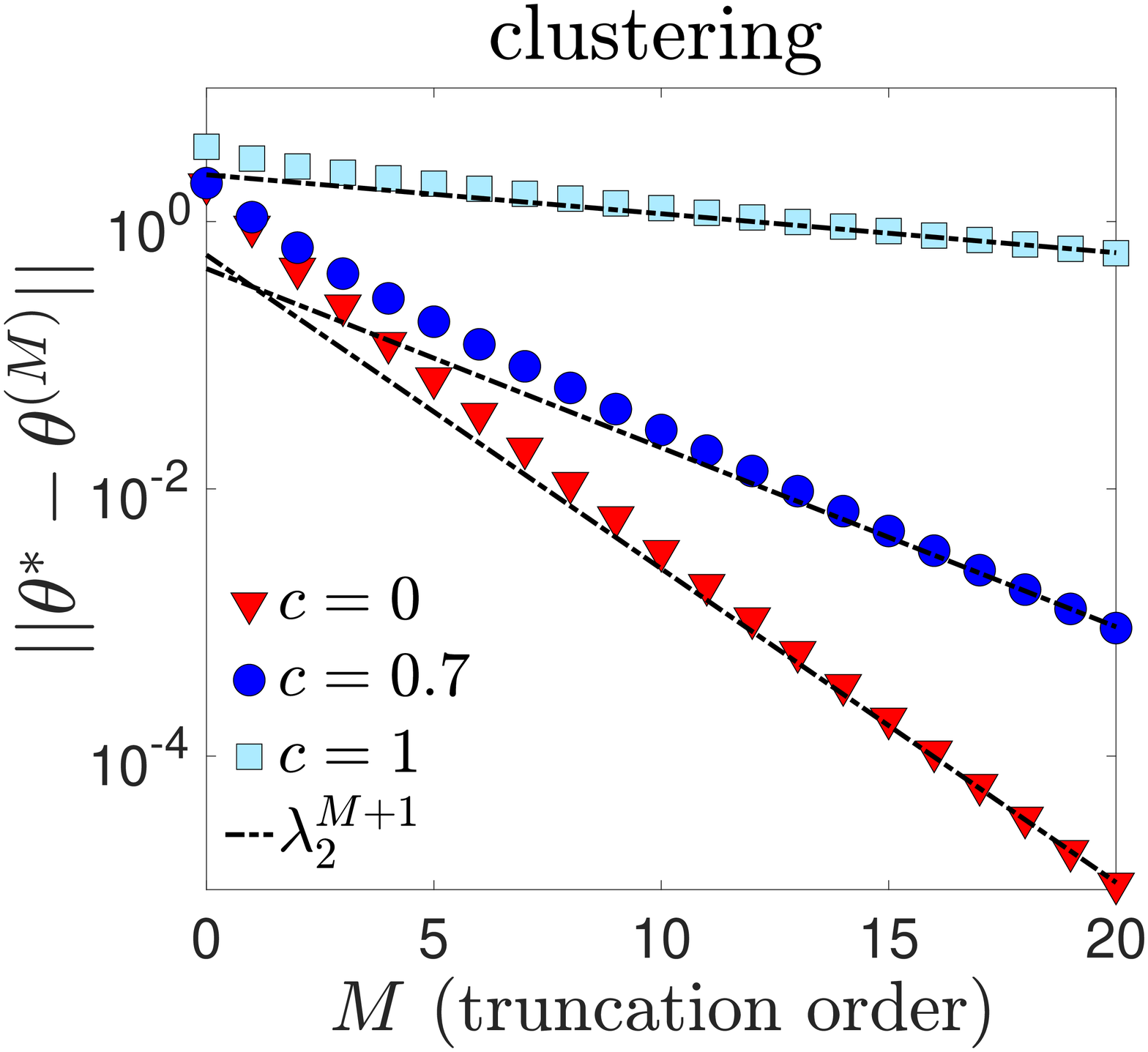}
\includegraphics[scale = 0.153]{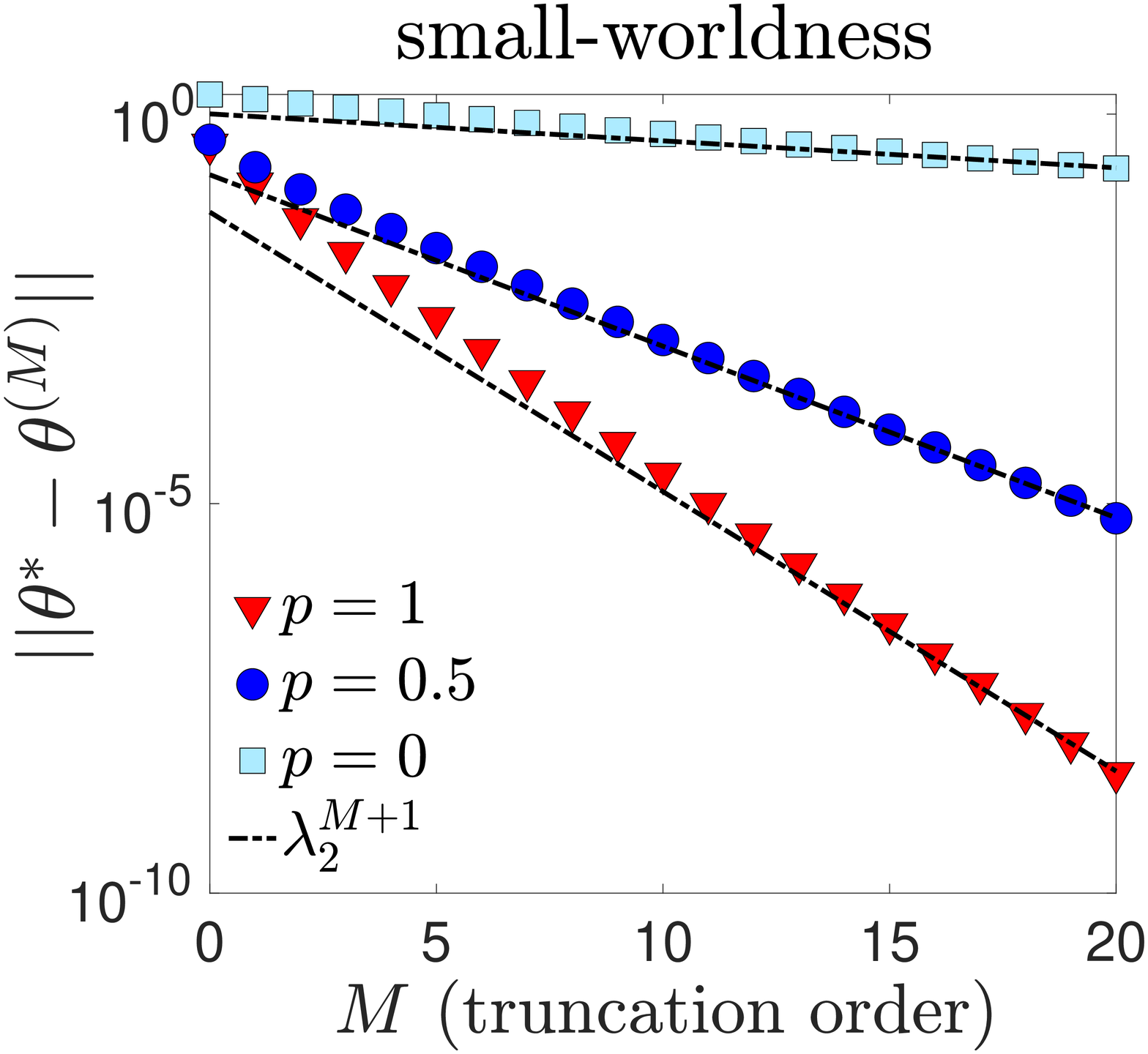}
\caption{\label{fig2} Error of the approximation depending on the truncation order for several interpolating network models, with fixed size $N = 1000$. We study the effect of density (varying the average degree), and fixing the mean degree at $\langle k \rangle = 10$, we also study the effect of modularity (interpolating from a random network to a 2-modules configuration), clustering (using the algorithm \cite{holme02} to generate SF networks with controllable clustering and the small-world model to interpolate between a ring and a random network \cite{watts98}. In all the scenarios, we observe that, for sufficiently high truncation order $M$, the error is dominated by $\lambda_2$, the second largest eigenvalue of the normalized adjacency matrix $B$. $\lambda_2$ increases with sparsity, modularity, and clustering as expected.}
\end{figure}

Turning now to a more computationally practical question, we investigate the rate of convergence of this expansion. Assumign for simplicity the case of an undirected network (so we may use Theorem~\ref{theorem1}), we consider the $M$-order approximation defined as $\bm{\theta}^{(M)} = \sum_{m = 0}^{M} ({D}^{-1}{A})^m {D}^{-1} \bm{\omega}$. Using Eq.~(\ref{equation12}), the error is given by 
\begin{equation}
\begin{split}
{D}^{\frac{1}{2}}(\bm{\theta}^* - \bm{\theta}^{(M)}) &=  \sum_{m = M+1}^\infty(\alpha_2\lambda_2^m \bm{v_2} + \cdots + \alpha_n \lambda_n^m \bm{v_n})\\
&=\frac{\alpha_2\bm{v_2}}{1-\lambda_2}\lambda_2^{M+1}+ \cdots + \frac{\alpha_N\bm{v_N}}{1-\lambda_N}\lambda_N^{M+1}.
\label{equation15}
\end{split}
\end{equation}
As the order $M$ of the approximation increases, the dominant term in Eq.~(\ref{equation15}) is that which corresponds to the second largest (in magnitude) eigenvalue $\lambda_2$ of the normalized adjacency matrix $B$. (Recall that the largest of eigenvalue of $B$ is $\lambda_2=1$.) Save for the unlikely scenario where $\bm{x}=D^{-1/2}\bm{\omega}$ is exactly orthogonal to $\bm{v}_2$, in which case $\alpha_2$ vanishes, then for large enough $M$ the mean square error will scale geometrically with the magnitude of this second largest eigenvalue, i.e.,
\begin{equation}
|| \bm{\theta}^* - \bm{\theta}^{(M)} || \sim |\lambda_2|^{M+1}.
\label{equation16}
\end{equation}
Therefore, the smaller $|\lambda_2|$ is, the quicker the approximation will converge. On the other hand, convergence will be slower for sparse networks with either strong modularity or clustering as well as strong bipartite structure, in which cases $\lambda_2$ tends to be close to $1$ and $-1$, respectively\cite{newman10,masuda17} and a larger number of terms are needed to obtain a desired level of accuracy. In Fig.~(\ref{fig2}) we plot the error of the truncated approximation depending on the truncation order for different values of the network density, modularity, clustering coefficient and small-worldness and compare against the theoretical scaling predicted by Eq.(\ref{equation16}) observing a perfect agreement for sufficiently large order $M$ in all the cases.

\section{Local approximation of synchronization}\label{section3}

Next we generate a local approximation from the geometric expansion presented above to describe the degree of synchronization of a system given by the Kuramoto order parameter\cite{kuramoto03}, $re^{i\psi} = N^{-1}\sum_j e^{i\theta_j}$. First, after the linearizing about a strongly synchronized state where all $|\theta_i|\ll1$ (note that an appropriate shift in initial conditions allows us to set the mean phase $\psi=0$) one obtains \cite{skardal14} $r \approx 1 - \frac{1}{2\sigma^2N^2}||\bm{\theta}^*||^2$. Note that the degree of synchronization increases, i.e., tends towards one, as the dispersion in the phases is reduced. By truncating Eq.~(\ref{equation9}) at the first-order term and neglecting the contribution of the shift to the mean, an approximation which is accurate for locally tree-like networks for which $|\lambda_j|\ll1$ for $j=2,\dots,N$, we have that 
\begin{equation}
\theta_i \approx \frac{\omega_i}{k_i} + \left \langle \frac{\omega}{k}\right \rangle_i^{(1)} 
\label{equation17}
\end{equation}
with $\langle \frac{\omega}{k}\rangle_i^{(1)} =  \frac{1}{k_i}\sum_{j = 1}^N a_{ij}\frac{\omega_j}{k_j}$ being the average contribution of the first neighbours arriving at node $i$. We can then directly write the order parameter as 
\begin{equation}
r \approx 1 - \frac{1}{2\sigma^2 N^2} \sum_{i = 1}^N \left(\frac{\omega_i}{k_i} + \left \langle \frac{\omega}{k}\right \rangle_i^{(1)} \right)^2.
\label{equation18}
\end{equation}

The local unfolding of synchronization dynamics from the geometric expansion that we use to write Eq.~(\ref{equation18}) allows us to gain analytical insight into the interplay between topology and dynamics that improves synchronization as well as understand several features that to date have only been investigated numerically\cite{skardal13,skardal14,skardal16a,taylor16,witthaut12,coletta16}, for instance \emph{i)} linear degree-frequency correlations reduce the absolute value of the phases at a given truncation order, which tend to reduce the overall dispersion (increasing synchronization) \cite{skardal13,skardal16a,gardenes11}, \emph{ii)} negative frequency-frequency correlations between connected neighbors tends to make the first order term of opposite sign (but smaller) to the local term, and this reduces the dispersion (increasing synchronization) \cite{skardal14,skardal16a}, \emph{iii)} synchronization is reduced when the network become sparser. Even if the loss of links is compensated by an increase of the weights, a sparse network will have more dispersion on the phases with respect to a denser one, because in the latter the higher-order terms sum over more neighbours and thus they vanish out more rapidly. This provides a clear mechanistic interpretation of weight localization phenomena \cite{skardal19}, and \emph{iv)} if frequencies are randomly allocated, $r \sim 1 -\langle k^{-2} \rangle$, where $\bm{k}$ are the strengths (or degrees of the nodes). The inverse second moment is larger for heterogeneous networks, thus homogeneity promotes synchronization in the linearized regime \cite{skardal19}. This is opposite to what occurs in the critical threshold, where $K_c \sim \langle k \rangle / \langle k^2 \rangle$, and homogeneity delays the critical threshold \cite{arenas08}. The interpretability of these effects emerge naturally from the local description of synchrony, and although the particularities of each phenomenon may require further analysis, the geometric unfolding allows to unveil the underlying mechanistic rules that control the interplay of structure and dynamics in synchronization.

Furthermore, the local approximation in Eq.~(\ref{equation18}) is useful in estimating the impact that removing or adding a link has on the degree of synchronization. In Ref.~\cite{taylor16} an approximation for the change in the order parameter $r$ is given for undirected networks using a perturbation analysis to approximate changes in the Laplacian eigenvalues, and in turn the change to the order parameter. Using the geometric expansion this problem can be addressed using only local information and directed networks may also be treated. First, we denote the change in the order parameter $r$ as a result of removing or adding a directed link coming from $q$ to $p$ as $\Delta r_{pq}$. Assume that degrees are sufficiently large and no frequency or degree correlations exist, it is straightforward to obtain from Eq.~(\ref{equation18}) that
\begin{equation}
\Delta r_{pq} \approx \mp \frac{1}{Nk_p}[\frac{\omega_p}{k_p}(\frac{\omega_p}{k_p}-\frac{\omega_q}{k_q})],
\label{equation19}
\end{equation}
where the $\mp$ sign accounts for removal/addition of links. First, the absolute value $|\Delta r_{pq}|$ will be larger if node $p$, i.e., the node with an incoming link from node $q$, has a large value of the ratio $\omega_p/k_p$ (i.e., large frequency and/or low degree) and the difference of phases (or ratios) between nodes $p$ and $q$ is large. Thus, the most important links in terms of their contribution to the degree of synchronization are those that connect nodes with large rations $|\omega/k|$. Subsequently, connecting very similar nodes and hose that point towards nodes with small $|\omega/k|$ contribute less to the degree of synchronization.

\begin{figure}[h]
\includegraphics[scale = 0.35]{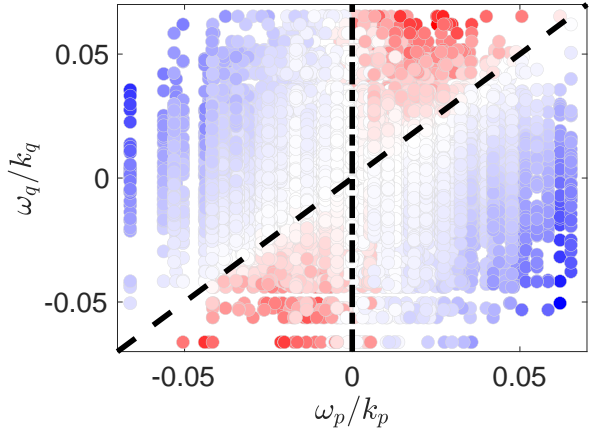}
\caption{\label{fig7}  Distribution of removed directed links $(p,q)$ in the phase-space $(\omega_p/k_p,\omega_q/k_q)$  for an Erd\"os-R\'enyi network with $N = 500$ and $\langle k \rangle = 50$, and gaussian $g(\omega)$ with $N(0,1)$. The color indicates whether if the removed link decreases (blue) or increases (red) the degree of synchrony, calculated exactly from Eq.(\ref{equation3}). Straight lines bound the region predicted by Eq.(\ref{equation20}) in the local approximation}
\end{figure}

Going one step further, using Eq.~(\ref{equation19}) we can predict which directed links will produce the counter-intuitive effect of increasing (decreasing) synchrony after its removal (addition). This effect, known as the Braess Paradox in the context of road traffic\cite{newman10}, has been studied for oscillatory networks\cite{witthaut12,coletta16,motter18} although the identification of these particular links relied on numerical schemes or expressions in terms of the spectral decomposition of $L$. Also, while these works study if the perturbations break the stability of the current state, here we assume that the perturbations in the links drive the system towards a new steady-state. In our formalism, we can directly impose $\Delta r_{pq} > 0$ for a link removal to obtain the condition
\begin{equation}
\frac{\omega_p}{k_p}(\frac{\omega_p}{k_p}-\frac{\omega_q}{k_q}) < 0.
\label{equation20}
\end{equation}
The condition in Eq.~(\ref{equation20}) describes two regions of the $(\omega_p/k_p,\omega_q/k_q)$-plane, namely the wedges $\omega_q/k_q>\omega_p/k_p$ for $\omega_p>0$ and $\omega_q/k_q<\omega_p/k_p$ for $\omega_p<0$. In terms of their area these wedges describe a quarter of $(\omega_p/k_p,\omega_q/k_q)$ space. In Fig.~(\ref{fig7}) we plot the resulting change $\Delta r_{pq}$ for each possible link removal in an Erd\"{o}s-R\'{e}nyi network of size $N=500$ with mean degree $\langle k\rangle=50$, color-coding the change so that positive (negative) changes are shaded more red (blue). We note that the positive changes fit well within the wedges predicted by our local theory, which are plotted in dashed black lines. In fact, approximately a quarter of directed links have the potential to increase synchronization after its removal as expected. While this phenomenon has been investigated in the context of identical oscillators\cite{arenas08,nishikawa06}, here the frequencies of the oscillators play a critical role in determining which directed links are harmful or redundant, and we have shown that the local approximation is sufficient to capture this phenomenon. 

\section{Conclusions}\label{section4}
In this paper we introduced a geometric expansion that expresses the exact solution of the linearized synchronization problem on a complex network as a Taylor series in contributions from increasingly further network neighborhoods. We proved that this series converges under relatively mild conditions, namely, that the network has a primitive adjacency matrix, and this method can be used for both directed and undirected networks. Moreover, the approximation error from using a truncated series scales geometrically with the second largest eigenvalue of the normalized adjacency matrix. We then applied these theoretical results to gain analytical and mechanistic insight on the optimal interplay of structure and dynamics on networks by deriving a local approximation for the degree of synchronization in the system. Furthermore, this approximation correctly identifies directed links may be removed to improve synchronization properties, a phenomenon known as the Braess paradox, as it captures the local interplay between network structure and dynamics.

Our results are of both theoretical and practical relevance for the study of synchronization on networks. The theory unfolds the stationary dynamics of the system in the real (geometric) domain of the nodes and it provides a complete understanding of the system from its local construction, unlike all previous results found in literature that solve many specific problems in terms of global spectral information. Also, we can use the theory to tackle important practical problems such as network inference\cite{timme07} and optimization\cite{skardal14,skardal16a,taylor16}, critical threshold estimation\cite{dorfler13}, control of synchronized states by local weight tuning or structure modifications\cite{skardal15b,arola18} and to predict how fluctuations in the weights or the frequencies propagate to the macroscopic synchronization \cite{arola20,skardal19}, just to name a few. Furthermore, our theory can be applied beyond the synchronization problem, since it finds utility in any forced diffusion system on networks that reduces to solving the linear system of Eq.~(\ref{equation5}). 

\begin{acknowledgments}
LA-F and AA acknowledges the Spanish MINECO, Grant No. FIS2015-71582-C2-1.
\end{acknowledgments}

\bibliography{aipsamp.bib}% Produces the bibliography via BibTeX.

\appendix

\section{Proof of Theorem 2}\label{appendixA}

In this Appendix we now present the proof of Theorem~\ref{theorem2}.
\begin{Proof}
Rather than making use of the symmetric normalized adjacency matrix, as in the Proof of Theorem~\ref{theorem1}, here we use the more classically stochastic matrix $D^{-1}A$. However, this matrix shares similar properties with its symmetric counterpart, namely, because $A$ is primitve, then so is $D^{-1}A$, and the Peron-Frobenius theorem guarantees similar eigenvalue properties, namely there is a single largest eigenvalue $\lambda_1$ that is real and larger in magnitude than all other eigenvalues, i.e., $\lambda_1>|\lambda_j|$ for $j=2,\dots,N$. Also, since $D^{-1}A$ is stochastic, $\lambda_1=1$ and $|\lambda_j|<1$ for $j=2,\dots,N$. On the other hand, the leading eigenvector is now given by the constant vector $\bm{v}_1\propto\bm{1}$. Importantly, since $D^{-1}A$ is not symmetric (even if $A$ is), the eigenvectors are not orthogonal to one another and cannot be used to form an orthonormal basis for $\mathbb{R}^N$. Nonetheless, we may use these eigenvalues as a (non-orthogonal) basis for $\mathbb{R}^N$ and uniquely expand the vector $\bm{x}=D^{-1}\bm{\omega}$ using this basis, specifically
\begin{align}
\bm{x}=\sum_{j=1}^N\alpha_j\bm{v}_j.\label{eq:appA:01}
\end{align}
We then look at each eigenmode $j=1,\dots,N$ of the term $\bm{\phi}_m$, namely $\bm{\phi}_m^{(j)}=\alpha_j(D^{-1}A)^m\bm{v}_j$ so that $\bm{\phi}_m=\sum_{j=1}^N\bm{\phi}_M^{(j)}$.
In terms of the full expression for $\bm{\phi}$ [given in Eq.~(\ref{equation13})] we then have that
\begin{equation}
\begin{split}
\bm{\phi}&=\sum\limits_{m=0}^\infty \sum_{j=1}^N\left(\bm{\phi}_m^{(j)}-\langle\bm{\phi}_m^{(j)}\rangle\right)\\
&=\sum_{j=1}^N\sum\limits_{m=0}^\infty \left(\bm{\phi}_m^{(j)}-\langle\bm{\phi}_m^{(j)}\rangle\right).
\label{eq:appA:02}
\end{split}
\end{equation}
We now treat the contribution of each eigenmode separately. We begin with the eigenmodes $j\ge2$ for which $|\lambda_j|<1$. First, we have that
\begin{align}
\bm{\phi}_m^{(j)}=\alpha_j(D^{-1}A)^m\bm{v}_j=\alpha_j\lambda_j^m\bm{v}_j,\label{eq:appA:03}
\end{align}
and 
\begin{align}
\langle\bm{\phi}_m^{(j)}\rangle=\langle\alpha_j\lambda_j^m\bm{v}_j\rangle=\alpha_j\lambda_j^m\langle\bm{v}_j\rangle,\label{eq:appA:04}
\end{align}
so together we have that 
\begin{equation}
\begin{split}
\sum\limits_{m=0}^\infty \left(\bm{\phi}_m^{(j)}-\langle\bm{\phi}_m^{(j)}\rangle\right)&=\sum_{m=0}^\infty\alpha_j\lambda_j^m\left(\bm{v}_j-\langle\bm{v}_j\rangle\right)\\
&=\frac{\alpha_j}{1-\lambda_j}\left(\bm{v}_j-\langle\bm{v}_j\rangle\right),
\label{eq:appA:05}
\end{split}
\end{equation}
i.e., each component converges for $j\ge2$.

To complete the proof we now show that the $j=1$ eigenmode, for which $\lambda_1=1$, converges. In fact, it turns out that this component has no contribution due to the shift of the mean. As in Eqs.~(\ref{eq:appA:03}) and (\ref{eq:appA:04}), we have that 
\begin{align}
\bm{\phi}_m^{(1)}=\alpha_1(D^{-1}A)^m\bm{v}_1=\alpha_1\lambda_1^m\bm{v}_1=\alpha_1\bm{v}_1,\label{eq:appA:06}
\end{align}
and  
\begin{align}
\langle\bm{\phi}_m^{(1)}\rangle=\langle\alpha_1\lambda_1^m\bm{v}_1\rangle=\alpha_1\lambda_1^m\langle\bm{v}_1\rangle=\alpha_1\langle\bm{v}_1\rangle,\label{eq:appA:07}
\end{align}
so
\begin{align}
\bm{\phi}_m^{(1)}-\langle\phi_m^{(1)}\rangle=\alpha_1\left(\bm{v}_1-\langle\bm{v}_1\rangle\right),\label{eq:10}
\end{align}
but since $\bm{v}_1\propto\bm{1}$, i.e., it's constant, we have that $\bm{v}_1=\langle\bm{v}_1\rangle$ and each term $\phi_1^{(m)}-\langle\phi_1^{(m)}\rangle$ vanishes, which completes the proof.\qed
\end{Proof}

\nocite{*}

\end{document}